\documentclass[useAMS,usenatbib]{mn2e}

\usepackage{times}
\usepackage{aas_macros}
\usepackage{graphicx}
\usepackage{amssymb}

\title[Old SMC star clusters]{On a possible origin for the lack of old star
clusters in the Small Magellanic Cloud}

\author[D. D. Carpintero, F. A. G\'omez and A. E. Piatti]
{D. D. Carpintero$^{1,2}$\thanks{E-mail: ddc@fcaglp.unlp.edu.ar},
F. A. G\'omez$^{3,4}$\thanks{E-mail: fgomez@pa.msu.edu}
and A. E. Piatti$^{5}$\thanks{E-mail: andres@mail.oac.uncor.edu}\\
$^{1}$Facultad de Ciencias Astron\'omicas y Geof\'\i sicas, Universidad Nacional de La Plata, Argentina\\
$^{2}$Instituto de Astrof\'\i sica de La Plata, UNLP-Conicet La Plata, Argentina\\
$^{3}$ Department of Physics and Astronomy, Michigan State University, East Lansing, MI 48824, USA\\
$^{4}$ Institute for Cyber-Enabled Research, Michigan State University, East Lansing, MI 48824, USA\\
$^{5}$Observatorio Astron\'omico, Universidad Nacional de C\'ordoba, Argentina}
\begin{document}

\date{}

\pagerange{\pageref{firstpage}--\pageref{lastpage}} \pubyear{}

\maketitle

\label{firstpage}

\begin{abstract}

  We  model the  dynamical  interaction between  the  Small and  Large
  Magellanic   Clouds   and   their  corresponding   stellar   cluster
  populations.   Our  goal is  to  explore  whether  the lack  of  old
  clusters ($\gtrsim 7$  Gyr) in the Small Magellanic  Cloud (SMC) can
  be the  result of  the capture of  clusters by the  Large Magellanic
  Cloud (LMC), as well as  their ejection due to the tidal interaction
  between the  two galaxies.  For this  purpose we perform  a suite of
  numerical  simulations probing a  wide range  of parameters  for the
  orbit  of  the  SMC  about  the  LMC.  We  find  that,  for  orbital
  eccentricities $e  \geq 0.4$, approximately  15 per cent of  the SMC
  clusters are captured  by the LMC.  In  addition, another 20 to
    50  per  cent  of its  clusters  are  ejected  into  the
    intergalactic medium.   In general, the  clusters lost by  the SMC
    are  the less  tightly bound  cluster population.   The  final LMC
    cluster distribution shows a spatial segregation between clusters
    that  originally   belonged  to  the  LMC  and   those  that  were
    captured from the SMC. Clusters  that originally belonged  to the SMC  are more
    likely  to be found  in the  outskirts of  the LMC.   Within this
  scenario it is possible to interpret the difference observed between
  the  star field  and cluster  SMC Age-Metallicity  Relationships for
  ages $\gtrsim 7$ Gyr.

\end{abstract}


\section{Introduction}

\citet{PG13} have claimed that the origin of the 15 known oldest Large
Magellanic Cloud (LMC) clusters still remains unexplained and
constitutes one of the most intriguing enigmas in our understanding of
the LMC formation and evolution. In addition, they mentioned that the
population of old Small Magellanic Cloud (SMC) clusters drastically
decreases beyond $\sim$ 7 Gyr and there is only one older than 10
Gyr. Furthermore, based on the statistics of catalogued and studied
clusters performed by \citet{P11} and the latest identified relatively
old SMC cluster \citep{P12}, a total of only six relatively old
clusters remain to be studied in this galaxy. From this result, it
arises the possibility of connecting the origin of the oldest LMC
cluster population to stripping events of ancient SMC star
clusters. Indeed, it is curious in this context that the oldest SMC
cluster is at the young and metal-rich extreme of the LMC globular
cluster distribution.

\citet{BKHVCK12} have showed that the observed irregular morphology and internal
kinematics of the Magellanic System (in gas and stars) are naturally explained
by interactions between the LMC and SMC, rather than gravitational interactions
with the Milky Way. They examined the gas and stellar kinematic centres of the
LMC; the warped LMC old stellar disk and bar; the gaseous arms stripped out of
the LMC by the SMC in the direction of the Magellanic Bridge; the stellar debris
from the SMC seen in the LMC disk field; etc., to strongly reinforce the
suspicions of \citet{dF72} that the interaction with the Milky Way is not
responsible for the LMC's morphology. Moreover, these conclusions provide
further support  that the Magellanic Clouds (MCs) are completing their first
infall to our Galaxy. 

As far as MC's star clusters are considered, \citet{BCBFCD04} proposed  that
differences in the birthplaces of both MCs and initial masses between the two
caused the LMC cluster age gap and the lack of old SMC clusters They suggested
that  the LMC/SMC were formed as different entities rather than as a binary 
protogalaxy in order to explain the difference in the MCs star cluster formation
history. On the other hand, \citet{PSGBC02} suggested that the SMC was formed
from the detachment of some part of the LMC containing gas and/or star clusters.
Notice that both works aimed at explaining the remarkably complementary
age-metallicity relationship observed in the MCs. 

In this Letter we trace for the first time the MCs star cluster dynamical
behaviour from numerical simulations in order to explore the possibility that
SMC globulars have been stripped out by the LMC. In Sect. \ref{numsi} we deal
with the computation of the star cluster orbits, whereas in Sect. \ref{res} we
discuss the probability of cluster capture in terms of different scenarios and
parameter values. In Sect. \ref{con} we summarise our results.

\section{Numerical simulations}
\label{numsi}

In order to investigate whether the above-mentioned hypothesis may be true, we
ran a series of numerical experiments. We chose $10^{10}$ M$_\odot$ as the unit
of mass, 1 kpc as the unit of distance, and the gravitational constant $G=1$.
This yields 4.7147 Myr as the unit of time and a unit of velocity equal to 207.4
km s$^{-1}$. The experiments consisted in following the evolution of the MCs in
mutual gravitational interaction on bound orbits, plus 100 point masses around
each cloud representing the clusters. The initial distance between both galaxies
was 100 kpc.

We simulated the MCs through Hernquist potentials \citep{H90}:
\begin{equation}
\Phi(r)=-\frac{GM}{r+a},
\end{equation}
where $M$ is the mass and $a$ is the scale-length of the model. We used $M_{\rm
L}=1.8\times 10^{11}$ M$_{\rm\sun}$ and $a_{\rm L}=21.4$ kpc for the LMC, and
$M_{\rm S}=2.1\times 10^{10}$ M$_{\rm\sun}$ and $a_{\rm S}=7.3$ kpc for the SMC
\citep{BKHVCK12}. 

We wanted to generate initial conditions for the bounded binary orbits of the
MCs with different eccentricities, in order to see whether exchanges of clusters
may be possible in different scenarios. However, since the MCs are not point
masses, a Keplerian eccentricity and its associated elliptic orbit are not
defined. We therefore defined the eccentricity as
\begin{equation}
e=\frac{r_{\rm a}-r_{\rm p}}{r_{\rm a}+r_{\rm p}},
\end{equation}
where $r_{\rm a}$  and $r_{\rm p}$ are the  pericentric and apocentric
distances of  the rosette orbit between the  two spherical potentials,
respectively. Then, to obtain an  orbit of one  galaxy about the
  other with  a given  $e$, we  first integrated a  test orbit  with a
  chosen initial separation  and with a guess on  the initial velocity
  (perpendicular to the  radius joining both MCs, i.e.,  we started at
  the apocentre of the orbit) and computed the resulting eccentricity.
  The initial separations considered were  60 and 100 kpc. By iterating
this process,  the initial velocity  corresponding to the  desired $e$
was found.  Finally,  the initial positions and velocities  of the MCs
thus  generated  were translated  to  the  centre  of mass  coordinate
system.

We then distributed 100 clusters of mass $10^5$ M$_{\rm\sun}$ around each
galaxy, with positions and velocities following the corresponding Hernquist
distribution function with isotropic velocity dispersion \citep{H90}:
\begin{eqnarray}
f({\bf x},{\bf v})=
\frac{M}{8\sqrt{2}\pi^3 a^3 w^3}\frac{1}{(1-s^2)^{5/2}} \nonumber \\
\times\left[ 3\arcsin s + s(1-s^2)^{1/2}(1-2s^2)(8s^4-8s^2-3)\right],
\label{frv}
\end{eqnarray}
where  $({\bf   x},{\bf  v})$   is  a  point   of  the   phase  space,
$w=(GM/a)^{1/2}$,          and         $s=(-E)^{1/2}/w$,         being
$E=\frac{1}{2}v^2+\Phi(r)$ the  energy per unit mass of  a cluster. In
the last equation  we have used the usual  notations $r=|{\bf x}|$ and
$v=|{\bf v}|$. The density profile turns out to be:
\begin{equation}
\rho(r)=\frac{M}{2\pi} \frac{a}{r(r+a)^3},
\end{equation}
though we added a cut  in this distribution at different fractions of the
  tidal radius, $R_{\rm t}$,  of each  galaxy (see details  below). The  tidal
  radius of the LMC (SMC) is 74 (26) kpc for an initial separation of 100
  kpc, and 46  (15) kpc for an initial separation of  60 kpc. The
  considered cut to the initial cluster distributions are $R_{t}$, $0.5 R_{\rm
    t}$ and, as a limiting case, a cut at 10 kpc for both galaxies. The initial
  radial extent  of a galaxy's cluster  population is still  a matter of
  debate.    Using    numerical   simulations,   \citet{B05}   \citep[see
  also][]{BY06} finds that this extent  can be approximated as a fraction
  of the half-mass radius of the host's dark matter halo. The half mass radius,
  $R_{\rm h}$, of the MCs can be estimated as $R_{\rm h} = 0.6
  R_{200}$ \citep[see][]{WGDRQ04}, where $R_{200}$ corresponds to the virial
  radius of the halo. Considering $R_{200} = 117.1~(57.1)$ kpc for
  the LMC (SMC) \citep{BKHVCK12}, we obtain a $R_{\rm h} = 70.3~(34.3)$ kpc. These
  values are consistent with the tidal radii obtained when
  an initial separation of 100 kpc between the clouds is considered. 

Since the MCs are fixed potentials, there is no dynamical friction in their
orbits. To make the simulations more realistic, we added this acceleration to the
velocities {\bf v} of the MCs in their orbits by using Chandrasekhar's dynamical
friction formula \citep{C43,BT08}:
\begin{equation}
\frac{{\rm d}{\bf v}_2}{{\rm d}t} =
-4 \pi G^2 M_2 \rho_1 \ln\Lambda \left[\int_0^{v_2} v^2 f_1(v) {\rm d}v \right]
\frac{{\bf v}_2}{v_2^3},
\label{cha}
\end{equation}
where the subindex 1 refers to the galaxy causing the friction, the subindex 2
refers to the galaxy being decelerated, {\bf v}$_2$ is the relative velocity of
the MCs, and $\Lambda$ is the Coulomb factor defined in our case as:
\begin{equation}
\Lambda = \frac{r v_2^2}{M_2(r)},
\end{equation}
where $r$ is the distance between the MC's centres, and $M_2(r)$ is the mass of
the galaxy 2 enclosed into $r$. The distribution function of the velocities of
the galaxy 1, $f_1(v)$, should be obtained from Eq. (\ref{frv}) by integrating
to all positions. Unfortunately, there is no closed form for the primitive, so
that the integral of Eq. (\ref{cha}) was approximated with:
\begin{equation}
\int_0^{v_2} v^2 f_1(v) {\rm d}v \simeq \frac{1}{6}\left( {\rm erf}(x)-
\frac{2}{\sqrt{\pi}} {\rm e}^{-x^2} \right),
\end{equation}
with $x=2v_2\sqrt{a_1 M_1}$. This approximation gives an error of less than 6
per cent at any $v_2$.

We used a tree code \citep{BH86,H87} to perform the experiments. But since we
considered both MCs as two particles not being point masses, we modified the
tree algorithm in order to self-consistently follow their evolutions. At each
time step, we first started by assigning a null mass to both galaxies. This
avoids the MCs being attracted to each other as Keplerian potentials, but
allowing the 200 clusters to be gravitationally influenced one to each other and
to accelerate their host galaxies. After computing all these accelerations, we
put the two Hernquist spheres centred at the positions where the galaxies were
found at that time, and added the accelerations produced by them on the clusters
and on each other as well. The system was then evolved in time, and a new time
step was taken.

\section{Results}
\label{res}

\begin{figure}
\includegraphics[width=\hsize]{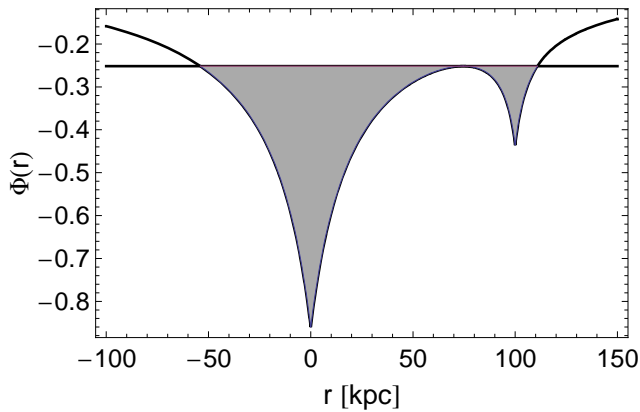}
\caption{Example of the total potential of the MCs along the line joining their
centres. The LMC and the SMC are here located at $r=0$ and $r=100$ kpc,
respectively; the saddle point determining the respective tidal radii is at
$r=74.53$ kpc. The straight line shows the value of the potential $\Phi_0$ at
the saddle point; the grey areas correspond to the regions of influence of each
galaxy, when the value of the equipotential $\Phi_0$ is taken as a reference.} 
\label{equip}
\end{figure}

For each  chosen $e$ value, we integrated  five different realisations of the 
cluster distributions --using different seeds  in each case--, in  order to be 
able to  estimate the  statistical uncertainties. During  the execution  of 
each  experiment, we  kept  track of  the membership of each cluster in  order
to detect any possible exchange between both galaxies.  To  this  purpose, we 
first computed the position of the  saddle point corresponding to the MCs, i.e.,
the  point at which the  acceleration due to  both galaxies is zero, as
exemplified in  Fig. \ref{equip}.  This point, which lies on the line joining
the centres of the MCs, approximately determines the instantaneous tidal radius
of each galaxy.  Note that this is an approximate   value  due   to  the   fact 
that   the  equipotential corresponding  to  the saddle  point  is  not 
spherical around  the centres of the MCs.

Let the potential of this saddle point at any time $t$ be $\Phi_0(t)$. We might
now decide whether a given cluster belongs to one or another galaxy by comparing
both its distance to the galaxy centres with the respective tidal radii, and its
energy (per unit mass) relative to the galaxies with $\Phi_0$. Fig.
\ref{equip} shows in grey the regions of membership when this criterium is
chosen. This procedure satisfactorily works when dealing with static potentials.
However, in our case, it has a serious drawback: since $\Phi_0(t)$ depends on
time, it could happen that some clusters have energies slightly below $\Phi_0$
at a given time, and slightly above $\Phi_0$ at the next time step, so that they
are "lost" due to the shifting of $\Phi_0$ without their dynamical status having
been significantly changed.

Since we are interested in real star cluster exchanges between the MCs, we
considered a star cluster to belong to a galaxy whenever the former is located
inside the tidal radius of the latter, irrespective of the energy. It might
happen that a star cluster --which has acquired too much energy-- be assigned to
a MC, even though it might be unbounded from both MCs. However, in such a case,
our bookkeeping would still result reasonable, since the star cluster will
quickly escape and go beyond any tidal radii.

\begin{figure}
\includegraphics[width=\hsize]{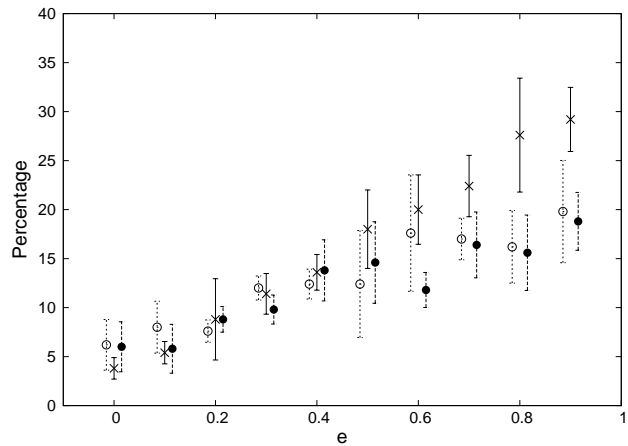}
\caption{Percentage of clusters that initially belonged to the SMC and, at the
time of the third apocentre, were captured by the LMC. Points and crosses refer
to experiments without and with dynamical friction, respectively. The latter
were slightly shifted to the right for a better readability. Also shown are the
results of experiments with dynamical friction but with initial separation
between the galaxies of only 60 kpc (open circles, slightly shifted to the left).
The error bars correspond to the standard deviation computed from the 5
experiments done for each case.} 
\label{cap}
\end{figure}

We  ran  each  experiment  until  the apocentre  following  the  third
pericentre  of the  orbit is  reached;  longer times  might imply  the
fusion  of   the  galaxies  or   the  tidal  disruption  of   the  SMC
\citep{BKHVCK12}.   For each  experiment, we  logged the  MC  to which
every  cluster  initially  belonged  to  and, at  the  final  time  of
integration, we  recomputed the  membership as described  above.  Fig.
\ref{cap} shows the percentage  of clusters that initially belonged to
the SMC and, at the time  of the third apocentre, were captured by the
LMC. For these  experiments we considered  a cut to  the initial
  radial extent  of the cluster  population equal to $R_{\rm  t}$. We
present the  results both with (filled circles)  and without (crosses)
dynamical friction.   As can be  seen, when the dynamical  friction is
taken into account there are far less captures than in the other case,
specially   at  large   eccentricities,  notwithstanding   the  closer
pericentres of the former case. This may be explained by the fact that
the closer a pericentre, the  smaller the time of high interaction, so
the MCs  interact more  strongly but during  a shorter period  of time
when the dynamical  friction is present. In any  case, captures appear
to be  a frequent, non-negligible  feature of the  interaction between
the  MCs.  Fig.   \ref{cap} also  shows the  outcome of  an additional
experiment  including  dynamical   friction,  but  with  the  galaxies
initially separated 60  kpc instead of 100 kpc  (open circles).  It is
clearly seen  that this parameter  doesn't play any important  role in
the  rate of  captures. In general,  when dynamical  friction is
  taken into account, approximately 15 per cent of the SMC clusters are
  captured by the LMC, for $e \geq 0.4$.

\begin{figure}
\includegraphics[width=\hsize]{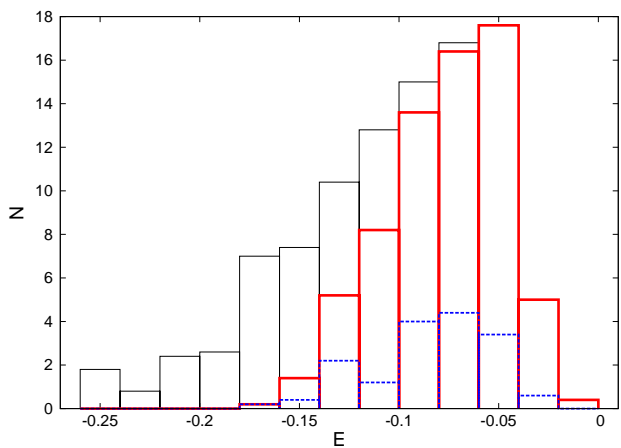}
\caption{Distribution  of  initial   binding  energies  of  the  SMC's
  clusters, for the case $e=0.7$ and with dynamical friction, averaged
  over the  five experiments. For  each bin, the clusters  captured by
  the LMC (lower areas), those  unbounded at the end of the experiment
  (middle areas), and  those that remained bounded to  the SMC (upper
  areas) are  also shown. The  bins correspond to spherical  shells of
  constant width.}
\label{histo}
\end{figure}

Fig. \ref{histo} shows, for  the case $e=0.7$ with dynamical friction,
the original distribution  of energies of the SMC's  clusters (we have
averaged the  values of  the five experiments  in order to  reduce the
noise). Each bin is subdivided into three parts, depending on the fate
of the  clusters. Those  captured by the  LMC correspond to  the lower
parts, limited  by dashed lines  (blue online); those which  were left
unbounded from  both galaxies are  piled above the latter,  limited by
thick lines (red online), and  those which remained bounded to the SMC
are represented by the upper parts,  piled on top. As can be seen, the
energy region  from which the clusters  are stripped out  from the SMC
correspond to  those less  bounded, i.e., those  farther out  from the
galaxy. Also, we may expect the intergalactic medium to be filled with
clusters lost from the SMC. Approximately 50 per cent of its clusters
are ejected. The experiments done with other eccentricities showed the
same trend.

It might  be suspected that  our choice of initial  conditions favours
the loss  of clusters  since, even though  they are  spatially located
inside their respective  galactic tidal radii at the  beginning of the
integrations,  some  of them  have  energies  above  the saddle  point
between  the MCs.  However, this  should not  be a  big  concern since
previous  tidal interactions with  other galaxies,  or even  the Milky
Way, are expected to  expand the original cluster radial distribution,
leaving some clusters on high energy orbits \citep{M86}.

Nevertheless,   we  ran   an  additional  set   of  experiments,
  considering  dynamical  friction, with  all  the clusters  initially
  placed inside  the tidal radii \emph{and} having  energies below the
  threshold of the  saddle point.  Note that this  condition imposes a
  more  centrally concentrated  radial distribution  of  clusters than
  before.   In what  follows, the  initial eccentricity  of  the SMC's
  orbit about  the LMC is fixed  at $e=0.7$, as this  is the preferred
  eccentricity  in  the  models  presented by  \citet{BKHVCK12}.   The
  experiments yielded $14\pm3$ per  cent of captures of SMC's clusters
  by the  LMC, whereas $20\pm 7$ and  $20\pm 6$ per cent  of LMC's and
  SMC's   clusters  were   thrown  into   the   intergalactic  medium,
  respectively.    Interestingly,  similar  percentages   of  captured
  clusters  were obtained when  different cuts  for the  initial radial
  extent  of the  cluster  populations were  considered.   For cuts  at
  $0.5R_{\rm t}$ and  10 kpc, we found that $17\pm 3$  and $15\pm 3$ per
  cent of SMC  clusters were captured by the  LMC, respectively. In the
  latter experiment  we found that $5 \pm  3$ and $29\pm 7$  per cent of the LMC
  and  SMC  clusters  were   ejected  into  the  intergalactic  medium,
  respectively.

We compared the final spatial distribution of the simulated
LMC's cluster population with the observed distribution showed by
\citet[][see their Fig. 10]{PGSG09}. Since our models can only be  
compared to the distribution of old halo-like pressure-supported  
stellar clusters, we
selected from the old cluster sample of \citet{PGSG09} only those with  
ages $\geq 7$ Gyr.
Note that this  is approximately the time at which the
  MCs    may   have    started   to    interact   with    each   other
  \citep{BKHVCK12,KVBAA13}.
Using these old clusters we find a mean
cluster projected distance with respect to the LMC's centre of approximately 5
kpc.
In order to compute the
projected distance in our models, we considered the $z=0$ plane
of our
simulations as the plane of the sky, and projected the simulated 3D distances
onto that plane.
Our models with
  an initial separation between the clouds of 100 kpc and a cut to the
  cluster's radial distribution at $R_{\rm  t}$ yielded a final mean LMC
  cluster's projected distance of  17 kpc. The large mean  distance obtained in
  these experiments can be attributed  to the large cut considered for
  the  initial   radial  distribution  of  clusters.    In  fact,  the
  experiments with an initial separation between the clouds of 60 kpc,
  and thus a much smaller  $R_{\rm t}$ (see Sec.~\ref{numsi}), yielded a
  projected mean  cluster  distance  of  10  kpc from  the  LMC's  centre.   For
  experiments  with cuts  at $0.5R_{\rm  t}$ and  10 kpc,  and initial
  separation of 100 kpc, we  obtained projected mean cluster distances of 16
  and 9  kpc, respectively.   It is interesting  to note that  in all
  cases  we found  a  spatial segregation  between  the clusters  that
  originally belonged to the LMC and those that were captured from the
  SMC. For instance, in the simulations  with an initial separation of 100
  kpc and an initial radial cut at 10 kpc, we found projected mean distances of 7
  and  21 kpc for the  original and  captured LMC  clusters, respectively.
  Although  less significant,  the same  was found  in  simulations with
  an initial separation of 60 kpc, where the resulting projected mean distances
were
  10 and 12 kpc, respectively. Thus, captured SMC clusters are more
  likely to be found in the outskirts of the LMC.

\section{Conclusions}
\label{con}

In this work we have explored the scenario in which the older and more
metal  poor  stellar  clusters  observed  in the  LMC  are,  at  least
partially, a sub-population of stellar clusters captured from the SMC.
For this purpose, we have  performed and analysed a suite of numerical
simulations of  the interaction of  MCs in isolation, probing  a large
number of different orbital  configurations and initial radial cluster
distributions.  The SMC was populated with a distribution of halo-like
pressure-supported  stellar clusters represented as  point masses.  In
all cases  the simulations  were allowed to  evolve for  three orbital
periods,  since longer interaction  times would  likely result  in the
complete disruption of the SMC \citep[][]{BKHVCK12}.

It has  been long known  that tidal interactions between  galaxies can result in
swapping and lost  of stellar clusters from their respective hosts \citep[see 
e.g.][]{M87}. However, this  is the first  time that this scenario  is explored
in  the context of the  interaction between the MCs.  Our  results show that,
for orbits of the  SMC about the LMC with eccentricities larger than $e = 0.4$,
a total of approximately $15$ per cent of  the SMC stellar clusters are 
captured by the  LMC. In  addition, a fraction of 20 to 50 per cent of  
its clusters are expelled into the intergalactic medium.  

Not surprisingly, only  the least bound clusters are  lost by the SMC.
As shown  by \citet{BCBFCD04}, the  tidal interaction between  the MCs
can induce the formation of  new stellar clusters in the inner regions
of  the galaxies,  especially in  the SMC  as it  is the  less massive
object and thus  more susceptible to the tidal  field.  However, these
newly formed clusters are more tightly bound to its host and thus more
likely to remain bound during the interaction. Note that the formation
of  clusters in  the LMC  should not  be triggered  until the  LMC has
interacted violently with the SMC, much later on, when the pericentric
passages  of  the SMC  get  sufficiently  close  to the  LMC's  centre
\citep[approximately  10 kpc apart,][]{BCBFCD04}. The  final LMC
  cluster distribution  shows an spatial  segregation between clusters
  that   originally  belonged  to   the  LMC   and  those   that  were
  captured.  Clusters that  originally belonged  to the  SMC  are more
  likely to be found in the outskirts of the LMC.

Recently,  \citet{PG13}   presented  a  detailed   comparison  of  the
Age-Metallicity Relationships (AMRs) obtained from field stars and stellar
clusters for both  the LMC and SMC.  Their  analysis provided evidence
for the formation of stars in the LMC between 5 and 12 Gyr, within the
well-known   cluster  age   gap,  with   almost   negligible  chemical
enrichment.  The cluster and field AMRs of the LMC show a satisfactory
match only for the last 3 Gyr,  while for older ages ( $> 11$ Gyr) the
cluster AMR  is a remarkable lower  envelope to the field  AMR. On the
other hand,  the field and  cluster AMRs observed  in the SMC  seem to
perfectly agree for the last 7 Gyr, time after which the population of
clusters becomes negligible. Interestingly,  for ages $ \gtrapprox 12$
Gyr the  mean metallicity of  the SMC's field stars  agrees remarkably
well with  that of  the LMC clusters.  This reinforces  our hypothesis
that, at least partially, the  older clusters observed in the LMC were
originally members of the SMC's cluster population.

Fully self-consistent hydrodynamical simulations of the interaction of
the MCs  and the  addition of tidal  field of  the Milky Way  would be
required to further explore this scenario. We defer this analysis to a
follow-up paper.

\section*{Acknowledgments}

The authors would like to thank the anonymous referee for the careful reading of
the manuscript and for the aptly suggestions and comments which allowed us to
improve the quality of our work. The  authors  also wish  to   thank  Gurtina 
Besla  for  providing  useful information about  the MCs'  orbital
configuration.  FAG  was supported through the  NSF Office  of
Cyberinfrastructure by  grant PHY-0941373, and  by  the Michigan  State 
University  Institute for  Cyber-Enabled Research (iCER). This work was
partially supported by the Argentinian institutions Universidad Nacional de La
Plata, Consejo Nacional de Investigaciones Cient\'\i ficas y T\'ecnicas, and
Agencia Nacional de Promoci\'on Cient\'\i fica y Tecnol\'ogica.

\bibliographystyle{mn2e.bst}
\bibliography{biblio.bib}

\bsp

\label{lastpage}

\end{document}